\documentclass[12pt]{article}
\usepackage{amsmath,amssymb,amsfonts,graphicx}

\begin{document}
\title{Polarization of the vacuum of quantized spinor field by a topological defect in two-dimensional space}%

\author{ Yurii A. Sitenko${}^{1}$ and Volodymyr M. Gorkavenko${}^{2}$\\
\it \small ${}^{1}$Bogolyubov Institute for Theoretical Physics,
 \it \small National Academy of Sciences of Ukraine,\\
 \it \small 14-b Metrologichna str., Kyiv 03143,
 Ukraine\\\phantom{bhh}\\
 \it \small ${}^{2}$Department of Physics, Taras Shevchenko National
 University of Kyiv,\\ \it \small 64 Volodymyrs'ka str., Kyiv
 01601, Ukraine}
 \date{}

\maketitle

\setcounter{page}{1}%

\begin{abstract}
Two-dimensional space with a topological defect is a transverse
section of three-dimensional space with the Abrikosov-Nielsen-Olesen
vortex, i.e. a gauge-flux-carrying tube which is impenetrable for
quantum matter. Charged spinor matter field is quantized in this
section with the most general mathematically admissible boundary
condition at the edge of the defect. We show that a current and a
magnetic field are induced in the vacuum. The dependence of the
results on boundary conditions is studied, and we find that the
requirement of finiteness of the total induced vacuum magnetic flux
removes an ambiguity in the choice of boundary conditions. The
differences between cases of massive and massless spinor matter are
discussed.
\end{abstract}

%\keywords{vacuum polarization; vortex; current; magnetic flux.}

\maketitle

\section{Introduction}

Topological phenomena are of great interest and importance because
of their universal nature connected with general properties of
space-time, on the one hand, and their numerous practical aspects,
on the other hand. In 1959, Aharonov and Bohm \cite{Aha} considered
the quantum-mechanical scattering of a charged particle on a
magnetic vortex and found an effect that does not depend on the
depth of penetration of the charged particle into the region of the
vortex flux. Thus, it was demonstrated for the first time that the
quantum-mechanical motion of charged particles can be affected by
the magnetic flux from the impenetrable for the particles region, if
the first homotopy group of the accessible to the particles region
is nontrivial. This effect which is alien to classical physics has a
great impact on the development of various fields in quantum
physics, ranging from particle physics and cosmology to condensed
matter and mesoscopic physics (see, e.g., reviews
\cite{Pes,Kri,Ton}). However even more important is that the
discovery of Aharonov and Bohm revealed a significance of topology
in the context of fundamental principles of quantum theory.

In 1957, Abrikosov \cite{Abr} discovered that a
magnetic vortex can be formed in the type-II superconductors, and
later this result was rederived in a more general context in
relativistic field theory \cite{NO}. Such string-like structures
denoted as the Abrikosov-Nielsen-Olesen (ANO) vortices arise as
topological defects in the aftermath of phase transitions with
spontaneous breakdown of continuous symmetries; the general
condition of the existence of these structures is that the first homotopy group of the group space of the broken symmetry group be nontrivial.

At present much attention is paid to the study of nonperturbative
effects in quantum systems, arising as a consequence of interaction
of quantized matter fields with a topologically nontrivial classical
field background. The ANO vortex is described classically in terms
of a spin-0 (Higgs) field which condenses and a spin-1 field
corresponding to the spontaneously broken gauge group; the former
field is coupled to the latter one in the minimal way with constant
${\tilde e}_{\rm cond}$. Single-valuedness of the condensate field
and finiteness of the vortex energy implement that the vortex flux
is related to ${\tilde e}_{\rm cond}$:
\begin{equation}\label{a11}
\Phi=\oint d \textbf{x} \textbf{V}(\textbf{x})=2\pi /{\tilde e}_{\rm
cond},
\end{equation}
where $\textbf{V}(\textbf{x})$ is the vector potential of the spin-1
gauge field, and the integral is over a path enclosing the vortex
tube once (natural units $\hbar=c=1$ are used). While considering
the effect of the ANO vortex on quantized matter fields, the
following circumstances should be taken into account. The quantized
matter field couples to a vector potential of the vortex-forming
gauge field in the minimal way with coupling constant $\tilde e$.
The ANO vortex is characterized by two cores: the one (where the
gauge symmetry is unbroken) has the transverse size of the order of
correlation length, and the other one (where the gauge field
strength is nonzero) has the transverse size of the order of
penetration depth. The value of the quotient of these length scales
is known as the Ginzburg-Landau parameter, and this value
distinguishes between superconductors of type-I and type-II (for a
review see, e.g., \cite{Hub}). Not going into details, we would like
to emphasize here that the phase with spontaneously broken symmetry,
where the matter field is quantized, exists in a spatial region
outside the vortex, and this region is not simply connected: its
first homotopy group is nontrivial. Hence, we anticipate in close
analogy with the Aharonov-Bohm effect that the ANO vortex  has no
effect on the surrounding matter in the framework of classical
theory, and such an effect, if exists, is of purely quantum nature.
Note also that, in view of the above, an issue of boundary
conditions for quantized matter field at the edge of the vortex
takes on special significance. The least restrictive, but still
physically acceptable, is the condition that ensures the
self-adjointness of the hamiltonian operator, see, e.g., \cite{Ree}.

The stress-energy tensor corresponding to the ANO vortex has
diagonal nonvanishing components only: $-T_{zz}=T_{00} > 0$,
$0 < -T_{rr} \ll T_{00}$, $0 < -r^{-2} \,\, T_{\varphi \varphi} \ll T_{00}$
(see, e.g., \cite{Gar}). The stress-energy tensor is a source of
gravity according to the Einstein--Hilbert equation
\begin{equation}\label{a12}
R_{\rho\rho'}-\frac 12 g_{\rho\rho'}R=8\pi GT_{\rho\rho'},
\end{equation}
where $R_{\rho\rho'}$ is the Ricci tensor,
$R=g^{\rho\rho'}R_{\rho\rho'}$ is the scalar curvature, $\c{G}$ is the gravitational constant; we use the notations adopted in \cite{Mis}. Taking the
trace over Lorentz indices in \eqref{a12}, one gets that the space-time
region of the vortex core is characterized by the positive scalar
curvature, $R>16\pi \c{G} T_{00}$, since $T_{00}$ is positive there. Space-time outside the vortex core is flat ($R=0$) but non-Minkowskian, with the metric given by squared length element
\begin{equation}\label{12}
    ds^2= - dt^2+dr^2+\nu^{-2} r^2 d\varphi^2+dz^2,
\end{equation}
where
\begin{equation}\label{13}
\nu=(1-4\c{G} M)^{-1},
\end{equation}
 $M$ is the linear density of mass stored in the core, which can be estimated to be of the order of the squared mass of the condensate field.
A transverse  ($z={\rm const}$) section of the outer space is a conical surface with the deficit angle equal to $8\pi\c{G} M$.

Quantum-field-theoretical models in $(2+1)$-dimensional space-time
exhibit a lot of interesting features, such as fermion number
fractionization, parity violation, and flavor symmetry breaking, for review see \cite{Dun,Mar}. A regular configuration
(i.e. a continuous in the whole function that can grow at most as
$O(|\textbf{x}-\textbf{x}_w|^{-2+\varepsilon})$ ($\varepsilon>0$) at
separate points) of magnetic field induces fermion number in the vacuum of quantized spinor matter field in two-dimensional space (surface) which is pierced by
the magnetic field strength lines; the density of fermion number is
proportional to the field strength, and the total fermion number is
proportional to the total field flux \cite{Nie}. The effect of a
singular configuration of magnetic field on the vacuum is quite
different; the point where the field strength pierces the surface is
punctured and the total vacuum fermion number which is induced on
the surface out of a puncture is periodic in the value of the total
flux of the singular field configuration. This was realized in a
rather general context in \cite{Si88,SiO,Si1}, where it was proven
for the first time that the flux through the nonaccessible for
quantized spinor matter field regions induces fermion number in its vacuum,
thus providing a manifestation of the Aharonov-Bohm effect (that is
characterized by the periodic dependence on the excluded magnetic
flux) \cite{Aha} in quantum field theory.

The case of the excluded magnetic flux is similar to the case of a
topological defect in the form of the ANO vortex, since in the
latter case the role of the magnetic field is played by the gauge
field corresponding to the spontaneously broken symmetry and the
vacuum of quantized spinor matter field exists out of the vortex core. As a
first step, one can neglect the transverse size of the vortex and
formally put the correlation length equal to zero. However, an issue
of the choice of boundary conditions even for the vortex with the
vanishing transverse size is of primary importance. This issue was
not touched upon in \cite{Si88,SiO,Si1}, it was elaborated later
with the use of the most general set of boundary conditions ensuring
the self-adjointness of the relevant Dirac hamiltonian operator.
Namely, all vacuum polarization effects which are induced by a
singular vortex in quantum spinor matter were obtained in
\cite{Si6,Si7,SiR,Si9a,Si9b} for the case of massive spinor and in
\cite{Si9c,Si9d,Si9e} for the case of massless spinor. It should be
noted that some vacuum polarization effects in the background of a
singular vortex were considered earlier in \cite{Gorn, Fle,Par} for
particular boundary conditions; however, the results in
\cite{Fle,Par} are actually erroneous, since periodicity in the value
of the flux of a singular vortex was overlooked in an endeavor to
imitate the results which are appropriate for the case of the
regular field configuration.

As a next step, one has to take into account conicity of space out
of the vortex. This task was considered in a number of papers, see
\cite{Fro,Dow,Gui,Mor,Bor,Iel,Srir,Spin}, for quantized both scalar
and spinor matter fields, sometimes incompletely and inconclusively as
regards to the case of massive matter. And the last step is to take
into account the nonvanishing correlation length, i.e. the transverse
size of the vortex. This task was considered in \cite{Bez1,Bel,Bez2}
for quantum spinor matter under a specific boundary condition, and
in \cite{Gor2,Gor3,Gor4,Gor1} for quantum scalar matter under the
Dirichlet boundary condition in space of arbitrary dimension.

The aim of the present work is to study the impact of boundary
condition of the most general form on the vacuum polarization
effects which are induced by the ANO vortex in quantum spinor matter
in $(2+1)$-dimensional space-time. Of primary interest are such
characteristics of the vacuum, as current, parity-violating
condensate and energy-momentum tensor\footnote{It should be noted
that current and energy-momentum tensor are the only vacuum
characteristics which are induced by the ANO vortex in quantum
scalar matter in space-time of arbitrary dimension, see
\cite{SiG,Si9,Si12}.}, since fermion number and angular momentum change
sign under the transition to the inequivalent irreducible
representation of the Dirac-Clifford algebra in $(2+1)$-dimensional
space-time.

\section{Current and magnetic field which are induced in the vacuum}

Postponing the consideration of parity-violating condensate and
energy-momentum tensor to subsequent publications, we start with
the induced vacuum current which is given by expression
\begin{equation}\label{17}
\textbf{j}(\textbf{x})= - \frac12\sum\hspace{-1.4em}\int \rm{sgn}(E)
\psi^\dag_E(\textbf{x}) \mbox{\boldmath $\alpha$}
\psi_E(\textbf{x}),
\end{equation}
where $\psi_{E}({\bf x})$ is the solution to the stationary Dirac
equation,
\begin{align}
& H \psi_{E}({\bf x})=E \psi_{E}({\bf x}),\nonumber \\
& H=-{\rm i} \mbox{\boldmath $\alpha$}\cdot \left(\mbox{\boldmath
$\partial$} - {\rm i}\tilde e\, \textbf{V}+\frac{{\rm i}}2
\mbox{\boldmath $\omega$}\right)+\beta m,\label{15}
\end{align}
$\textbf{V}({\bf x})$ and $\mbox{\boldmath $\omega$}({\bf x})$ are
the bundle and spin connections, symbol \mbox{$\displaystyle \sum\hspace{-1.4em}\int $} denotes
summation over the discrete part and integration over the continuous part of the energy spectrum, and $\rm{sgn}(u)$ is the sign function
($\rm{sgn}(u)=\pm 1$ at $u \gtrless 0$). As a consequence of the
Maxwell equation,
\begin{equation}\label{18}
\mbox{\boldmath $\partial$}\times \textbf{B}_{\rm I}(\textbf{x}) =
e\, \textbf{j}(\textbf{x}),
\end{equation}
the magnetic field strength, $\textbf{B}_{\rm I}(\textbf{x})$, is
also induced in the vacuum; here the electromagnetic coupling
constant, $e$, differs in general from $\tilde e$. The total flux of
the induced vacuum magnetic field is
\begin{equation}\label{19}
\Phi_I=\int d \mbox{\boldmath $\sigma$} \cdot \textbf{B}_{\rm
I}(\textbf{x}).
\end{equation}
In the background of the ANO vortex, the only one component of the
bundle and spin connections is nonvanishing:
\begin{equation}\label{113}
    V_\varphi=\frac{\Phi}{2\pi}, \quad w_\varphi={\rm
    i}\frac{\nu-1}r\, \alpha_\varphi \alpha_r,
\end{equation}
and the Dirac hamiltonian operator takes form
\begin{equation}\label{114}
H=-{\rm i}
\left[\alpha^r\left(\partial_r+\frac{1-\nu}{2r}\right)+\alpha^\varphi\left(\partial_\varphi-{\rm
i}\frac{\tilde e \Phi}{2\pi}\right)\right]+\beta m,
\end{equation}
where
\begin{align}
& \alpha^r=\alpha_r=\left(\begin{array}{cc}
                    0& {\rm i}  e^{-{\rm i}\varphi}\\
                   - {\rm i}  e^{{\rm i}\varphi} &0
                     \end{array}\right),\nonumber\\
& \alpha^\varphi=\frac{\nu}r\left(\begin{array}{cc}
                    0&  e^{-{\rm i}\varphi}\\
                     e^{{\rm i}\varphi} &0
                     \end{array}\right),\quad
\alpha_\varphi=\frac{r^2}{\nu^2}\,\alpha^\varphi.\label{115}
\end{align}
Decomposing function $\psi_E(\textbf{x})$ as
\begin{equation}\label{116}
\psi_E(\textbf{x}) = \sum_{n \in \mathbb{Z}}
                   \left(\begin{array}{c}
                   f_n(r,E )e^{ {\rm i} n\varphi} \\
                   g_n(r,E )e^{ {\rm i} (n+1)\varphi}
                    \end{array}\right)
\end{equation}
($\mathbb{Z}$   is the set of integer numbers), we present the Dirac
equation as a system of two first-order differential equations for
radial functions:
\begin{align}
&\left[-\partial_r + \frac{\nu \left(n-n_{\rm c} - F  +
\frac12\right)-\frac12}r\right] f_n =(E+m) g_n, \nonumber
\\
& \left[\partial_r + \frac{\nu \left(n-n_{\rm c} - F   +
\frac12\right)+\frac12}r \right] g_n =(E-m) f_n,\label{117}
\end{align}
where
\begin{equation}\label{118}
n_{\rm c}=\left[\!\left| \frac{\tilde e \Phi}{2\pi}
\right|\!\right],\quad F = \left\{\!\!\left| \frac{\tilde e
\Phi}{2\pi}  \right|\!\!\right\},
\end{equation}
$\left[\!\left| u \right|\!\right]$ is the integer part of quantity
$u$ (i.e. the integer which is less than or equal $u$), and $
\left\{\!\!\left|  u \right|\!\!\right\} = u - \left[\!\left| u
\right|\!\right]$ is the fractional part of quantity $u$, $ 0\leq
\left\{\!\!\left| u \right|\!\!\right\}<1 $. Using \eqref{115} and
\eqref{116}, one gets $j_r =0$, and the
 only component of the induced vacuum current,
\begin{equation}\label{119}
j_\varphi(r) = - \frac r\nu \sum\hspace{-1.4em}\int \sum_{n \in
\mathbb{Z}} {\rm sgn} (E) f_n(r,E) g_n(r,E),
\end{equation}
is independent of the angular variable. The induced vacuum magnetic
field strength is directed along the vortex axis,
\begin{equation}\label{120}
B_{\rm I}(r) = e \nu \int\limits_r^\infty \frac{dr'}{r'} \,
j_\varphi(r'),
\end{equation}
with total flux
\begin{equation}\label{121}
\Phi_{\rm I} = \frac{2\pi}\nu \int\limits_{r_0}^\infty dr\, r B_{\rm
I}(r),
\end{equation}
where it is assumed without a loss of generality that the vortex
core has the form of a tube of radius $r_0$.

We prove that the most general boundary condition ensuring the
self-adjointness of operator $H$ \eqref{114} is
\begin{equation}\label{71}
    (I-{\rm i} \beta\alpha^r\, e^{-{\rm i} \theta
    \alpha^r})\left.\psi\right|_{r=r_0}=0,
\end{equation}

\noindent where $\theta$ is the self-adjoint extension parameter.
This condition is also the most general one ensuring the absence of
the matter flux across the vortex core edge, i.e. the confinement of
the matter field to the region out of  the vortex core. Imposing
boundary condition \eqref{71} on the solution to the Dirac equation,
$\psi_E(\textbf{x})$ \eqref{116}, we obtain the condition for the
modes:
\begin{equation}\label{310}
\cos\left(\frac{\theta}{2}+\frac{\pi}4 \right)
f_n(r_0,E)=-\sin\left(\frac{\theta}{2}+\frac{\pi}4
\right)g_n(r_0,E).
\end{equation}
Using the explicit form of the modes satisfying \eqref{117} and
\eqref{310}, we derive the analytic expressions for the induced
vacuum current, $j_\varphi(r)$ \eqref{119}, and the induced vacuum
magnetic field, $B_{\rm I}(r)$ \eqref{120}, in the case of $\nu \geq 1$ and $0 < F < 1$, and in the case of $\frac12 \leq \nu < 1$ and
$\frac12\left(\frac1\nu -1\right) < F < \frac12\left(3-\frac1\nu
\right)$. The results can be
presented in the form
\begin{equation}\label{414}
j_\varphi(r)\!=\! j_\varphi^{(a)}(r) + j_\varphi^{(b)}(r; r_0),\,
B_{\rm I}(r)\!=\!B^{(a)}_{\rm I}(r)+B^{(b)}_{\rm I}(r; r_0),
\end{equation}
where all dependence on $r_0$ is contained in $j_\varphi^{(b)}$ and
$B^{(b)}_{\rm I}$, moreover,
\begin{equation}\label{415}
\lim_{r_0\rightarrow 0}j_\varphi^{(b)}(r; r_0)=0,
\quad \lim_{r_0\rightarrow 0}B_{\rm I}^{(b)}(r; r_0)=0.
\end{equation}
The crucial point is the behavior of $j_\varphi^{(b)}$ and
$B^{(b)}_{\rm I}$ at $r \rightarrow r_0$. If
\begin{equation}\label{416}
\lim_{r \rightarrow r_0}j^{(b)}_\varphi(r; r_0) \, (r-r_0)^2 = 0 \end{equation}
and, consequently,
\begin{equation}\label{417}
\lim_{r \rightarrow r_0}B_{\rm I}^{(b)}(r; r_0) \, (r-r_0) = 0,
\end{equation}
then flux $\Phi_{\rm I}$ \eqref{121} is finite. A careful numerical
analysis reveals that condition \eqref{416} is fulfilled in cases
$\theta=0$ and $\theta=\pi$ only. The case of $F=1/2$ needs a
special comment, because of oddness in $\theta$ in this case.
Whereas the current and, consequently, the induced magnetic field
with its flux vanish at $\theta=0$, they are nonvanishing and
discontinuous in $\theta$ at $\theta=\pi$. Namely, we obtain
\begin{multline}\label{524c}
\left.\Phi_{\rm
I}\right|_{F=1/2,\,\theta=\pi_{\pm}} \\
= \pm\frac{e}{8m}\,e^{2mr_0}\left[\Gamma(2,2mr_0)-4m^2r_0^2\Gamma(0,2mr_0)\right],
\end{multline}
where
$$\Gamma(z,u)=\int\limits_u^\infty dy\,y^{z-1}{\rm e}^{-y} $$
is the incomplete gamma-function; in particular,
\begin{equation}\label{524d}
\lim_{r_0\rightarrow0}\left.\Phi_{\rm
I}\right|_{F=1/2,\,\theta=\pi_\pm}=\pm\frac{e}{8m}.
\end{equation}
In the case of $F\neq 1/2$ continuity in $\theta$ is maintained, and
we obtain the following representation for the induced vacuum
magnetic flux:
\begin{equation}\label{527}
\left. \Phi_{\rm I}\right|_{\theta = \frac{\pi}{2} \mp
\frac{\pi}{2}} =\left. \Phi_{\rm I}^{(a)}\right|_{\theta = \frac{\pi}{2} \mp
\frac{\pi}{2}} +\left. \Phi_{\rm I}^{(b)}\right|_{\theta = \frac{\pi}{2} \mp
\frac{\pi}{2}}, \quad F \neq 1/2,
\end{equation}
where
\begin{multline}\label{528}
\left. \Phi_{\rm I}^{(a)}\right|_{\theta = \frac{\pi}{2} \mp
\frac{\pi}{2}} \\
= \frac{e}{4\nu m}\Biggl\{
\sum_{p=1}^{\left[\!\left| {\nu}/2 \right|\!\right]}
\exp[-2mr_0\sin(p\pi/\nu)]
\,\frac{\sin[(2F-1)p\pi]}{\sin^3(p\pi/\nu)} \Biggr. \\
\Biggl. - \frac{\nu}{4N} \left(-1\right)^{N}\sin\left(2NF \pi \right) {\rm e}^{-2mr_0}
\, \delta_{\nu, \, 2N} \Biggr\} \\
+{\rm
sgn}\left(F-\frac12\right)\frac{e}{8\pi m} \int\limits_0^\infty
\frac{du}{\cosh^3(u/2)}\,{\rm e}^{-2mr_0
\cosh(u/2)} \\
\times \Biggl\{ \cos\left[\nu\left(F-\frac12\right)\pi)\right]
\cosh\left[\nu\left(\left|F-\frac12\right|-1\right)u\right]\Biggr.
\\
\Biggl. - \cos\left[\nu\left(\left|F-\frac12\right|-1\right)\pi\right]\cosh\left[\nu\left(F-\frac12\right)u\right]\Biggr\} \\\times \vphantom{\frac12} \left[\cosh(\nu u)-\cos(\nu \pi)\right]^{-1},
\end{multline}
\begin{multline}\label{529}
\left. \Phi_{\rm I}^{(b)}\right|_{\theta = \frac{\pi}{2} \mp
\frac{\pi}{2}} \\
= \frac{e}{(4\pi)^2}  r_0\int\limits_{m
r_0}^\infty
 \frac{dv\, v}{\sqrt{v^2-m^2 r_0^2}}\\
 \times\Biggl\{\frac12\Biggl[\,{\rm
sgn}\left(F-\frac12\right)\Biggl(C^{(\pm)}_{\frac12+\nu\left(F-\frac12\right)}(v)+
C^{(\pm)}_{\frac12-\nu\left(F-\frac12\right)}(v)\Biggr)\Biggr.\Biggr.\\
\Biggl. + C^{(\pm)}_{\frac12+\nu\left(F-\frac12\right)}(v)-
C^{(\pm)}_{\frac12-\nu\left(F-\frac12\right)}(v)\Biggr]
D_{\frac12+\nu\left|F-\frac12\right|}(v)\\
 + \sum_{l=1}^\infty \left.[C^{(\pm)}_{\nu
\left(l+F-\frac12\right)+\frac12}(v) D_{\nu
\left(l+F-\frac12\right)+\frac12}(v) \right. \\
\left. \left. - C^{(\pm)}_{\nu \left(l-F+\frac12\right)+\frac12}(v)
D_{\nu \left(l-F+\frac12\right)+\frac12}(v) \right]\right\},
\end{multline}
\begin{multline}\label{526}
C^{(\pm)}_\rho(v)=\left\{v I_\rho(v)K_\rho(v)\pm\,
mr_0\left[I_\rho(v)K_{\rho-1}(v)\right.\right. \\
\left.\left. - I_{\rho-1}(v)K_\rho(v)\right]  -
vI_{\rho-1}(v)K_{\rho-1}(v)
\right\} \\
\times \left[v K^2_\rho(v)\pm 2mr_0K_\rho(v)K_{\rho-1}(v) +
vK^2_{\rho-1}(v)\right]^{-1},
\end{multline}
and
\begin{multline}\label{530}
D_{\rho}(v) = \rho K_\rho^2(v)-(\rho-1)K_{\rho+1}(v)K_{\rho-1}(v)
\\+ v \left[ K_{\rho}(v)\frac{d}{d \rho} K_{\rho-1}(v) -
K_{\rho-1}(v)\frac{d}{d \rho} K_{\rho}(v) \right].
\end{multline}
In particular,
\begin{multline}\label{531}
\left. \lim_{r_0\rightarrow 0} \Phi_{\rm I}\right|_{\theta =
\frac{\pi}{2} \mp \frac{\pi}{2}} = -\frac{e}{6m} \left[F -\frac12 -
\frac12{\rm
sgn}\left(F-\frac12\right)\right] \\
\times \Biggl\{\frac34 -\nu^2 \left[\frac14 - \left|F
-\frac12\right| - F(1-F)\right] \Biggr\}, \quad F \neq 1/2.
\end{multline}

The case of massless quantized spinor field is characterized by certain peculiarities. First, there is invariance under
transformation $\theta\rightarrow \pi - \theta$. Thus the results
are continuous in $\theta$, and their values at $\theta=0$ and
$\theta=\pi$ coincide, in particular,
\begin{equation}\label{94}
 \left.j_\varphi(r)\right|_{F=\frac12, \,
\theta=0}=\left.j_\varphi(r)\right|_{F=\frac12, \, \theta=\pi}=0
\end{equation}
and
\begin{equation}\label{95}
\left.B_{\rm I}(r)\right|_{F=\frac12, \, \theta=0}=\left.B_{\rm
I}(r)\right|_{F=\frac12, \, \theta=\pi}=0.
\end{equation}
Secondly, instead of exponential decrease, $j_\varphi$ and $B_{\rm I}$  decrease as  $r^{-1}$ at large distances from the ANO vortex.
Consequently, flux $\Phi_{\rm I}$, see \eqref{121}, is given by an
integral which is linearly divergent at $r \rightarrow \infty$.
Therefore, we have no choice but to introduce cutoff  $r_{\rm max} >
r$ and the restricted flux,
\begin{equation}\label{96}
\Phi_{{\rm I} (r_{\rm max})}=\frac{2\pi}{\nu}
\int\limits_{r_0}^{r_{\rm max}} dr\,r B_{\rm I}(r).
\end{equation}
As follows from our numerical analysis of the integrand in
\eqref{96} near the lower limit of integration, relation \eqref{416}
is fulfilled and flux $\Phi_{{\rm I} (r_{\rm max})}$ \eqref{96} is
finite at $\theta=\frac{\pi}{2} \mp \frac{\pi}{2}$ only. We get
immediately:
\begin{equation}\label{911}
 \left.\Phi_{{\rm I} (r_{\rm max})}\right|_{\theta = \frac{\pi}{2} \mp \frac{\pi}{2}}
= 0, \quad F = 1/2.
\end{equation}
As to $F \neq 1/2$, although we obtain the analytic expression for
$\left.\Phi_{{\rm I} (r_{\rm max})}\right|_{\theta = \frac{\pi}{2}
\mp \frac{\pi}{2}}$ for arbitrary $r_{\rm max} > r_0$, the
physically sensible case is that of $r_{\rm max} \gg r_0$. Retaining
only terms which are maximal in the latter case, we get the expression for the flux:
\begin{multline}\label{915}
\left.\Phi_{{\rm I} (r_{\rm max})}\right|_{\theta = \frac{\pi}{2}
\mp \frac{\pi}{2}} = \frac{e \, r_{\rm max}}{4\nu}
\left\{\sum_{p=1}^{\left[\!\left| {\nu}/2
\right|\!\right]} \frac{\sin[(2F-1)p\pi]}{\sin^2(p\pi/\nu)} \right.\\
\left. - \frac{\nu}{4N} (-1)^{N}\sin(2N F \pi) \, \delta_{\nu, \, 2N} \right\} \\
+\, {\rm sgn}\left(F-\frac12\right)
\frac{e \, r_{\rm max}}{8\pi}\int\limits_0^\infty \frac{du}{\cosh^2(u/2)} \\
 \times \left\{\cos\left[\nu\left(F-\frac12\right)\pi)\right]
\cosh\left[\nu\left(\left|F-\frac12\right|-1\right)u\right]\right.\\-
\left.
\cos\left[\nu\left(\left|F-\frac12\right|-1\right)\pi\right]\cosh\left[\nu\left(F-\frac12\right)u\right]\right\}\\
\times \vphantom{frac12} \left[\cosh(\nu u)-\cos(\nu \pi)\right]^{-1}
 + O\left(e \, r_0\right),  \quad  F \neq 1/2,
\end{multline}
and the relation between the current and the magnetic field:
\begin{multline}\label{916}
\nu e \left.j_\varphi(r)\right|_{\theta = \frac{\pi}{2} \mp
\frac{\pi}{2}}=\frac {r_{\rm max}}{r_{\rm max} - r} \left.B_{\rm
I}(r)\right|_{\theta = \frac{\pi}{2} \mp
\frac{\pi}{2}}\\=\frac{\nu}{\pi r_{\rm max} r}\left.\Phi_{{\rm I}
(r_{\rm max})}\right|_{\theta = \frac{\pi}{2} \mp \frac{\pi}{2}},
\quad  r \gg r_0.
\end{multline}
In particular, we get in the case of $\nu=1$:
\begin{multline}\label{add1}
\left.\Phi_{{\rm I} (r_{\rm max})}\right|_{\nu=1,\,\theta=\frac{\pi}2\mp
\frac{\pi}{2}} \\
= \frac{e}4\, r_{\rm max} \tan(F\pi) \left|F-\frac12\right|
\left(\left|F-\frac12\right|-1\right)+O(er_0)
\end{multline}
and
\begin{multline}\label{add2}
\left. e j_\varphi(r)\right|_{\nu=1,\,\theta=\frac{\pi}2\mp \frac{\pi}{2}} \\
=\frac{r_{\rm max}}{r_{\rm max}-r}\,\left. B_{I}(r)\right|_{\nu=1,\,\theta=\frac{\pi}2\mp \frac{\pi}{2}} \\
= \frac{e}{4\pi r} \tan(F\pi) \left|F-\frac12\right|
\left(\left|F-\frac12\right|-1\right), \,\, r\gg r_0.
\end{multline}
The last relation for the current was first obtained in \cite{Si9e}
(see (10.6) in this reference where the definition of the current
differs by extra $r^{-1}$). Note a discontinuity at $F=1/2$, which
is independent of $\nu$,
\begin{equation}\label{add3}
\lim_{F\rightarrow (1/2)_{\pm}} \left. e j_\varphi(r)\right|_{\theta \neq \pm
\frac{\pi}{2}}=\pm\frac{e}{4\pi^2
r},\,\, r\gg r_0.
\end{equation}
This is distinct from the case of quantized scalar field under
the Dirichlet boundary condition, when the current is continuous and
vanishing at $F=1/2$ \cite{Si9,SiB1,SiB2}, see the appropriate
expression from these references at $m=0$ and $\nu=1$:
\begin{equation}\label{add4}
\left. e j_\varphi(r)\right|_{scalar,\,Dirichlet}=-\frac{e}{4\pi
r}\tan(F\pi) \left(F-\frac12\right)^2.
\end{equation}

\section{Discussion and conclusion}

The effects of conicity, which are characterized by the value of the
deficit angle, $8\pi\c{G} M$, are negligible for the ANO vortices in
ordinary superconductors, since constant $\c{G}$ is of order of the
Planck length squared and quantity $M$ is of order of the inverse
correlation length squared. However topological defects of the type
of ANO vortices also arise in another field -- in cosmology and high
energy physics, where they attained the name of cosmic strings
\cite{Vil,Hin}.  Cosmic strings with $8\pi\c{G} M \sim 1$ are
definitely ruled out by astrophysical observations, but there
remains a room for cosmic strings with $8\pi\c{G} M \sim 10^{-6}$
and less (see, e.g., \cite{Bat}), although the direct evidence for
their existence is lacking.

A recent development in material science also provides an unexpected link
between condensed matter and high energy physics, which is caused to a large
extent by the experimental discovery of graphene -- a two-dimensional crystalline allotrope formed by a monolayer of carbon atoms \cite{Ge}. A single topological defect (disclination) warps a sheet of graphene, rolling it into a nanocone which is similar to the transverse section of a spatial region out of a cosmic
string; carbon nanocones with deficit angles equal to $N_d \pi/3$ ($N_d$ is a possible number of sectors which are removed from the hexagonal lattice: $N_d = 1, 2, 3, 4, 5$,
i.e. $\nu = \frac 65, \frac 32, 2, 3, 6$) were observed experimentally, see \cite{Nae}
and references therein. Moreover, theory also predicts saddle-like nanocones with the
deficit angle taking negative values unbounded from below (sectors can be added: $N_d =-1,-2,-3,-4,-5,-6,...,-\infty$,
i.e. $\nu = \frac 67, \frac 34, \frac 23, \frac 35, \frac {6}{11}, \frac 12,..., 0$), which can be regarded as corresponding
to cosmic strings with negative mass density. Note that nanoconical structures may arise
as well in a diverse set of condensed matter systems known as the two-dimensional Dirac
materials, ranging from honeycomb crystalline allotropes (silicene and germanene \cite{Cah},
phosphorene \cite{Liu}) to high-temperature cuprate superconductors \cite{Tsu} and topological
insulators \cite{Qi}.

Since the transverse size of the ANO vortex is related to the
correlation length, its nonvanishing value, $r_0$, should be taken
into account. We have considered the current and the magnetic field
which are induced in the vacuum of quantized spinor field in the
case of $\nu \geq 1$ and $0 < F < 1$, and in the case of $\frac12
\leq \nu < 1$ and $\frac12\left(\frac1\nu -1\right) < F <
\frac12\left(3-\frac1\nu \right)$. The dependence of these
characteristics of the vacuum on boundary conditions ensuring the
impenetrability of the vortex core is analyzed, and we find that the
demand of finiteness of the total induced vacuum magnetic flux
removes an ambiguity in the choice of boundary conditions. The case
of massless quantized spinor field requires an introduction of the
maximal size of the system, $r_{\rm max}$. We discover that, for
physically sensible values $r_{\rm max}\gg r_0$, the vacuum
polarization effects in this case, in distinction from the case of
massive quantized spinor field, are independent of the transverse
size of the ANO vortex. Due to this distinction, the results in the
massless case are discontinuous at $F=1/2$ with a jump which is
independent of $\nu$, whereas the results in the massive case at
$\theta=0$ are continuous in $F$ and vanishing at $F=1/2$, as long
as the transverse size of the vortex is nonvanishing.

\vskip3mm \textit{The work was presented at the XI International Bolyai-Gauss-Lobachevsky (BGL-2019) Conference on Non-Euclidean, Non-Commutative Geometry and Quantum Physics, May 19-24, Kyiv, Ukraine. Yu.A.S. acknowledges a support from the National Academy of Sciences of Ukraine (Project No.01172U000237), from the Program of Fundamental Research of the Department of Physics and Astronomy of the National Academy of Sciences of Ukraine (Project No.0117U000240),
and from the ICTP -- SEENET-MTP project NT-03 `Cosmology - Classical
and Quantum Challenges'.}


\begin{thebibliography}{0}    %for 1 digit

\bibitem{Aha}
Y. Aharonov and D. Bohm. Significance of Electromagnetic Potentials
in the Quantum Theory. \textit{Phys. Rev.} {\bf 115}, 485 (1959).

\bibitem{Pes}%
M. Peshkin and A. Tonomura. {\it The Aharonov--Bohm Effect} (Berlin:
Springer-Verlag, 1989).

\bibitem{Kri}%
I.V. Krive  and A.S. Rozhavsky. Non-traditional Aharonov-Bohm
effects in condensed matter. {\it Int. J. Mod. Phys. B} \textbf{6}
1255 (1992).

\bibitem{Ton}%
A. Tonomura.  The AB effect and its expanding applications.
\textit{J. Phys. A: Math. Theor.} \textbf{43} (2010) 354021.

\bibitem{Abr}
A.A. Abrikosov. On the Magnetic Properties of Superconductors of the
Second Group. \textit{Sov. Phys.-JETP} \textbf{5}, 1174 (1957).

\bibitem{NO}
H.B. Nielsen and P. Olesen. Vortex-line models for dual strings.
\textit{Nucl. Phys. B} {\bf 61}, 45 (1973).

\bibitem{Hub}%
R.P. Huebener. {\it Magnetic Flux Structure in Superconductors}
(Berlin: Springer-Verlag, 1979).

\bibitem{Ree}%
M. Reed and B. Simon. {\it Methods of Modern Mathematical Physics
II. Fourier Analysis, Self-Adjointness} (Academic Press, New York,
1975).

\bibitem{Gar}%
D. Garfinkle. General relativistic strings. {\it Phys. Rev. D}
\textbf{32} 1323 (1985).

\bibitem{Mis}%
C.W. Misner, K.S. Thorne and J.A. Wheeler.  {\it Gravitation} (San
Francisco: W H Freeman, 1973).

\bibitem{Dun}
G.V. Dunne. {\it Topological Aspects of Low Dimensional Systems} (Springer, Berlin, 1999).

\bibitem{Mar}
E.C. Marino. {\it Quantum Field Theory Approach to Condensed Matter Physics} (Cambridge: Cambridge University Press, 2017).

\bibitem{Nie} A.J. Niemi and G.W. Semenoff. Fermion number fractionization in quantum field theory. \textit{Phys. Rep.} \textbf{135}, 99
(1984).

\bibitem{Si88} Yu.A. Sitenko. On the electron charge fractionization in magnetic field with boundaries present. \textit{Sov. J. Nucl. Phys.} \textbf{47}, 184
(1988).

\bibitem{SiO} Yu.A. Sitenko. Electron-charge fractionization on surfaces of various geometry in an external magnetic field. \textit{Nucl. Phys. B} \textbf{342}, 655 (1990).

\bibitem{Si1} Yu.A. Sitenko. Geometry of the base manifold and fermion number fractionization. \textit{Phys. Lett. B} \textbf{253}, 138 (1991).

\bibitem{Si6} Yu.A. Sitenko. Self-adjointness of the Dirac hamiltonian and fermion number fractionization in the background of a singular magnetic vortex. \textit{Phys. Lett. B} \textbf{387},  334 (1996).

\bibitem{Si7} Yu.A. Sitenko. Self-adjointness of the Dirac Hamiltonian and vacuum quantum numbers induced by a singular external field. \textit{Phys. Atom. Nucl.} \textbf{60},  2102 (1997); (E) \textbf{62}, 1084 (1999) .

\bibitem{SiR} Yu.A. Sitenko and D.G. Rakityansky. Quantum Numbers of the Theta Vacuum in (2+1)-Dimensional Spinor Electrodynamics: Charge and Magnetic Flux. \textit{Phys. Atom. Nucl.} \textbf{60},  1497 (1997).

\bibitem{Si9a}  Yu.A. Sitenko. Effects of fermion vacuum polarization by a singular magnetic vortex: Zeta function and energy. \textit{Phys. Atom. Nucl.} \textbf{62},  1056 (1999).

\bibitem{Si9b}
Yu.A. Sitenko. Polarization of a fermion vacuum by a singular
magnetic vortex: Spin and angular momentum. \textit{Phys. Atom.
Nucl.} \textbf{62},  1767 (1999).

\bibitem{Si9c} Yu.A. Sitenko. Induced vacuum condensates in the background of a singular magnetic vortex in 2+1-dimensional space-time. \textit{Phys. Rev. D} \textbf{60}, 125017
(1999).

\bibitem{Si9d} Yu.A. Sitenko. Chiral symmetry breaking as a consequence of nontrivial spatial topology. \textit{Mod. Phys. Lett. A} \textbf{14}, 701
(1999).

\bibitem{Si9e} Yu.A. Sitenko. Self-adjointness of the two-dimensional massless Dirac Hamiltonian and vacuum polarization effects in the background of a singular magnetic vortex.
 \textit{Annals Phys.} \textbf{282}, 167 (2000).

\bibitem{Gorn} P. Gornicki. Aharonov-Bohm effect and vacuum polarization. \textit{Annals Phys.} \textbf{202}, 271 (1990).

\bibitem{Fle} E.G. Flekkoy and J.M. Leinaas. Vacuum currents around a magnetic flux string. \textit{Int. J. Mod. Phys. A}
\textbf{6}, 5327 (1991).

\bibitem{Par}  R.R. Parwani and A.S. Goldhaber. Decoupling in (2 + 1)-dimensional QED? \textit{Nucl. Phys. B} \textbf{359}, 483 (1991).

\bibitem{Fro}%
V.P. Frolov   and E.M. Serebriany. Vacuum polarization in the
gravitational field of a cosmic string.  {\it Phys. Rev. D}
\textbf{35} 3779 (1987).

\bibitem{Dow}%
J.S. Dowker. Vacuum averages for arbitrary spin around a cosmic
string.  {\it Phys. Rev. D} \textbf{36} 3742 (1987).

\bibitem{Gui}%
M.E.X. Guimaraes  and B. Linet. Scalar Green's functions in an
Euclidean space with a conical-type line singularity.  {\it Commun. Math. Phys.} \textbf{165} 297 (1994).

\bibitem{Mor}%
E.S. Moreira. Massive quantum fields in a conical background. {\it Nucl. Phys. B} \textbf{451} 365 (1995).

\bibitem{Bor} M. Bordag, K. Kirsten, and S. Dowker. Heat-Kernels and functional determinants on the generalized cone. \textit{Commun. Math. Phys.} \textbf{182}, 371 (1996).

\bibitem{Iel}%
D. Iellici. Massive scalar field near a cosmic string.  {\it Class. Quantum Grav.} \textbf{14} 3287 (1997).

\bibitem{Srir}%
L. Sriramkumar. Fluctuations in the current and energy densities around a magnetic flux carrying cosmic string. {\it Class. Quantum Grav.} \textbf{18} 1015 (2001).

\bibitem{Spin}  J. Spinelly and E.R. Bezerra de Mello.
Spinor Green function in higher-dimensional cosmic string space-time in the presence of magnetic flux. \textit{J. High Energy Phys.} \textbf{09}, 005 (2008).

\bibitem{Bez1}
E.R. Bezerra de Mello, V. Bezerra, A.A. Saharian, and V.M.
Bardeghyan. Fermionic current densities induced by magnetic ux in a
conical space with a circular boundary. \textit{Phys. Rev. D} {\bf 82}, 085033 (2010).

\bibitem{Bel}
S. Bellucci, E.R. Bezerra de Mello, and A.A. Saharian.Fermionic
condensate in a conical space with a circular boundary and magnetic flux.  \textit{Phys. Rev. D} {\bf 83}, 085017 (2011).

\bibitem{Bez2}
E.R. Bezerra de Mello, F. Moraes, and A.A. Saharian. Fermionic
Casimir densities in a conical space with a circular boundary and magnetic flux. \textit{Phys. Rev. D} {\bf 85}, 045016 (2012).

\bibitem{Gor2}
V.M. Gorkavenko, Yu.A. Sitenko, and O.B. Stepanov. Polarization of the vacuum of a quantized scalar field by an impenetrable magnetic vortex of finite thickness. \textit{J. Phys. A: Math. Theor.} {\bf 43}, 175401 (2010).

\bibitem{Gor3}
V.M. Gorkavenko, Yu.A. Sitenko, and O.B. Stepanov. Vacuum energy induced by an impenetrable flux tube of finite radius. \textit{Int. J. Mod. Phys. A} {\bf 26}, 3889 (2011).

\bibitem{Gor4}
V.M. Gorkavenko, Yu.A. Sitenko, and O.B. Stepanov. Casimir energy and force induced by an impenetrable flux tube of finite radius. \textit{Int. J. Mod. Phys. A} {\bf 28}, 1350161 (2013).

\bibitem{Gor1}
V.M. Gorkavenko, I.V. Ivanchenko, and Yu.A. Sitenko. Induced vacuum current and magnetic field in the background of a vortex. \textit{Int. J. Mod. Phys. A} {\bf 31}, 1650017 (2016).

\bibitem{SiG} Yu.A. Sitenko and V.M. Gorkavenko. Induced vacuum energy-momentum tensor in the background of a (d-2)-brane in (d+1)-dimensional space-time. \textit{Phys. Rev. D} \textbf{67}, 085015 (2003).

\bibitem{Si9} Yu.A. Sitenko and N.D. Vlasii. Induced vacuum current and magnetic field in the background of a cosmic string.  \textit{Class. Quantum Grav.}
\textbf{26}, 195009 (2009).

\bibitem{Si12} Yu.A. Sitenko and N.D. Vlasii. Induced vacuum energy-momentum tensor in the background of a cosmic string.
 \textit{Class. Quantum Grav.} \textbf{29}, 095002 (2012).

\bibitem{SiB1}%
Yu.A. Sitenko and A.Yu. Babansky. The Casimir-Aharonov–Bohm
effect? {\it Mod. Phys. Lett. A} \textbf{13}, 379 (1998).

\bibitem{SiB2}%
Yu.A. Sitenko   and A.Yu. Babansky. Effects of boson-vacuum
polarization by a singular magnetic vortex. {\it Phys. At. Nucl.} \textbf{61}, 1594 (1998).

\bibitem{Vil}%
A. Vilenkin and E.P.S Shellard. {\it Cosmic Strings and Other
Topological Defects} (Cambridge: Cambridge University Press, 1994).

\bibitem{Hin}%
M.B. Hindmarsh  and T.W.B. Kibble. Cosmic strings. {\it Rep. Prog. Phys.} \textbf{58} 477, (1995).


\bibitem{Bat}%
R.A. Battye, B. Garbrecht, A. Moss, and H. Stoica. Constraints on brane inflation and cosmic strings. \textit{J. Cosmol. Astropart. Phys. JCAP} \textbf{0801}, 020 (2008).


\bibitem{Ge}
A.K. Geim and K.S. Novoselov. The rise of graphene. \textit{Nature Mater. } {\bf 6}, 183 (2007).

\bibitem{Nae}
S.N. Naess, A. Elgsaeetter, G. Helgesen, and K.D. Knudsen. Carbon nanocones: Wall structure and morphology. \textit{Sci. Technol. Adv. Mat.} {\bf 10}, 065002 (2009).

\bibitem{Cah}
 S. Cahangirov, M. Topsakal, E. Akturk, H. Sahin, and
 S. Ciraci. Two- and One-Dimensional Honeycomb Structures of Silicon and Germanium. \textit{Phys. Rev. Lett.} {\bf 102}, 236804 (2009).

\bibitem{Liu}
H. Liu, A.T. Neal, Z. Zhu, Z. Luo, X. Xu, D. Tomanek, and P.D. Ye. Phosphorene: an unexplored 2D semiconductor with a high hole mobility. {\it ACS NANO} {\bf 8}, 4033 (2014).

\bibitem{Tsu}
C.C. Tsuei and J.R. Kirtley. Pairing symmetry in cuprate
superconductors. \textit{Rev. Mod. Phys.}  {\bf 72}, 969 (2000).

\bibitem{Qi}
X.L. Qi and S.C. Zhang. Topological insulators and superconductors. \textit{Rev. Mod. Phys.} {\bf 83}, 1057 (2011).


\end{thebibliography}
\end{document}